\documentclass[preprint2]{aastex}

\shorttitle{Massive Black Holes in Clusters}
\shortauthors{Baumgardt, Makino, Ebisuzaki}

\begin{document}

\title{Massive Black Holes in Star Clusters. I. Equal-mass clusters}
\author{Holger Baumgardt\altaffilmark{1},
        Junichiro Makino\altaffilmark{2},
	Toshikazu Ebisuzaki\altaffilmark{1}}

\altaffiltext{1}{
        Astrophysical Computing Center, RIKEN, 2-1 Hirosawa, Wako-shi,
        Saitama 351-0198, Japan}

\altaffiltext{2}{
        Department of Astronomy, University of Tokyo, 7-3-1 Hongo,
        Bunkyo-ku,Tokyo 113-0033, Japan}

\begin{abstract}
In this paper we report results of collisional $N$-body simulations of the dynamical 
evolution of equal-mass star clusters containing a massive central black hole.
Each cluster is composed of between 5,000 to 180,000 stars together with a central
black hole which contains between 0.2\% to 10\% of the total cluster mass.

We find that for large enough black hole masses, the central density follows a power-law 
distribution with slope $\rho \sim r^{-1.75}$ inside the radius of influence of the black hole, in 
agreement with predictions from earlier Fokker Planck and Monte Carlo models. 
The tidal disruption rate of stars is within a factor of two of that derived in previous
studies. It seems impossible 
to grow an intermediate-mass black hole (IMBH) from a $M \le 100 M_\odot$ progenitor
in a globular cluster by the tidal disruption of stars, although $M = 10^3 M_\odot$
IMBHs can double their mass within a Hubble time in dense globular clusters. The same is true
for the supermassive black hole at the centre of the Milky Way.

Black holes in star clusters will feed mainly on stars tightly bound to them and
the re-population of these stars causes the clusters to expand, reversing core-collapse
without the need for dynamically active binaries. Close encounters of stars 
in the central cusp also lead to an increased mass loss rate in the form of high-velocity 
stars escaping from the cluster. 
A companion paper will extend these results to the multi-mass case.
\end{abstract}


\newcommand{\msun}{M_{\odot}}

\section{Introduction}

Theoretical studies of the dynamics of massive black holes in dense stellar systems started in the
1960s to explain the central activity and luminosities of quasars. Since then,
the dynamics of a massive body in the centre of a stellar system has been the focus of a large 
number of theoretical studies, starting with classic papers by \citet{Peebles1972}, \citet{BahcallWolf1976,
BahcallWolf1977}, \citet{FrankRees1976} and \citet{CohnKulsrud1978}.

The problem is of great importance to astrophysics since
the centres of the Milky Way and other nearby galaxies contain black holes of $10^6$ to $10^9 
M_\odot$ \citep{KormendyGebhardt2001}.
In addition, smaller sized black holes of a few thousand solar masses might exist in globular clusters
\citep{Gerssenetal2002, Gerssenetal2003, Gebhardtetal2002, PortegiesZwartetal2004},
although the evidence for them is still controversial (Baumgardt et al. 2003ab).
A massive black hole in a galactic nucleus or a star cluster is a potential source of gravitational radiation
due to its high mass and the fact that it will frequently undergo close encounters with other stars and black holes
if the density of the surrounding
system of stars is large enough. It would therefore be a prime target for the forthcoming
generation of ground and space-based gravitational wave detectors. 

Intermediate-mass black holes of several 100 to several 1000 $M_\odot$
could also be the explanation for the ultraluminous X-ray sources observed in external galaxies 
\citep{PortegiesZwartetal2004, DiStefanoetal2004} and could provide the missing link
between the stellar mass black
holes formed as the end products of the stellar evolution of massive stars
and the $10^6 - 10^9 M_\odot$ sized black holes found in galactic centres \citep{Ebisuzakietal2001}.

\citet{BahcallWolf1976} showed that an equi\-librium-flow solution for stars into the gravitational well around a
black hole exists and predicted that
the stellar density will follow a power-law distribution $\rho=r^{-\alpha}$ with exponent $\alpha=1.75$. While Peebles (1972) found
a steeper slope, Monte Carlo simulations by \citet{CohnKulsrud1978} and \citet{MarchantShapiro1980} 
confirmed the results of \citet{BahcallWolf1976}. Due to the high stellar densities around the black hole,
tidal disruption of stars is important for the evolution of the system. 
\citet{FrankRees1976}
and \citet{LightmanShapiro1977} found that stars on highly eccentric orbits dominate the consumption
rate since stars drift faster in angular momentum space than in energy space. \citet{FrankRees1976} 
derived analytic formulae for the disruption rates which were later confirmed 
by Monte Carlo simulations \citep{MarchantShapiro1980,
DuncanShapiro1983}. In the latter paper the authors also showed that black holes in globular clusters preferentially
disrupt stars most tightly bound to the black hole while in galactic nuclei the disruption process
should be dominated by stars not bound to the black hole. 
The gas lost from disrupted stars is either accreted onto the central black holes or lost in a
stellar wind due to radiation drag, with the first process dominating for high enough black
hole masses \citep{Davidetal1987}.

Stellar collisions could also be an important process,
although for the small mass black holes expected in globular clusters,
\citet{CohnKulsrud1978} found that the collision rate of stars is about 30 times smaller than their consumption
rate by the black hole. Similarly, \citet{Murphyetal1991} performed multi-mass Fokker-Planck
calculations and found that in low-density galactic nuclei stellar disruptions happen more 
often than stellar collisions. 

Recently, \citet{Amaro-Seoaneetal2004} followed the evolution of a system of equal-mass stars with a central black hole
by means of an anisotropic gaseous model and found a strong black hole growth accompanied by the expansion of the cluster. 
An overall cluster expansion due to the inward drift and tidal disruption of stars
was also predicted by \citet{Shapiro1977}. 

Although various aspects of the dynamical evolution of black holes in dense stellar systems have been 
studied in the literature, nobody has tried a self-consistent direct $N$-body simulation of the
growth of the central BH. Moreover, simulations with approximate
methods such as Monte-Carlo or Fokker-Planck methods have been applied
only to idealised systems. For example, with the exception of \citet{BahcallWolf1977} and 
\citet{FreitagBenz2002}, most simulations so far considered only
single-mass systems, and ignored stellar evolution. Thus, though
such studies are useful to gain physical understanding of the problem,
they do not tell us much about the actual behaviour of star clusters
with central black holes. Can an IMBH grow from a smaller mass seed black hole by accreting
nearby stars? What will star clusters with an IMBH look like? Do they have cusps
in surface luminosity? In the present and companion papers, we address
these issues.
Our focus will be mainly on the dynamics of star clusters containing black holes. This is because
direct $N$-body simulations cannot be done for systems containing more than a few times $10^5$ stars.
In addition, simplifying assumptions like a fixed black hole at the cluster centre 
are most likely
violated for stellar systems containing black holes of only a few hundred to a few thousand times the
mass of a star. 

\section{Description of the runs}

We simulated the evolution of star clusters containing between $N = 5,000$ and $N=178,800$ stars
using the collisional Aarseth $N$-body code NBODY4 \citep{Aarseth1999} on the GRAPE6 boards of Tokyo 
University \citep{Makinoetal2003}. All clusters were treated as isolated and followed King profiles initially.
At the start of the calculation, the massive black holes were at rest at the cluster centres.
We set up the initial models so that the systems were in dynamical
equilibrium and the density profiles were the same as the corresponding
King models without the central black holes. To achieve this, we
calculated the new distribution function from the King model density
profile and the potential obtained by the original King potential
plus the BH potential, and generated positions and velocities of the stars
using this new distribution function. Note that it is impossible to setup an
equilibrium model with isotropic velocity dispersion, which has a flat
core with finite central density around a black hole \citep{Tremaineetal1994,
NakanoMakino1999}. Thus, our initial model is not in exact
dynamical equilibrium, and the 0.5\%  mass shell shows contraction of
about 10\% in case of a 5\% BH mass. The effect is however much
smaller than the initial contraction of the cluster with a central black hole 
in Fig. \ref{exp1}, which is caused by
the development of an $\alpha=-1.75$ cusp due to thermal evolution and can therefore
be neglected.

Two series of simulations were produced. In the first series we followed the  
evolution of equal-mass star clusters with central black holes.
These simulations were designed to identify the relevant
physical mechanisms and compare our results
with theoretical estimates and results reported in the literature.
In the second series of simulations we studied the dynamics of black holes in galactic globular 
clusters. We simulated multi-mass clusters 
containing between $16,384 < N < 131,072$ 
stars and a central black hole of $M_{BH} = 1000 M_\odot$. The results of these runs will be 
reported in a companion paper \citep{Baumgardtetal2004}.

In the present paper, stars were assumed tidally disrupted if their distance to the central black hole was 
smaller than 
a fixed disruption distance which was either $r_t=10^{-7}$, $10^{-8}$ or $10^{-9}$ in $N$-body units.
For solar type stars, the radius against tidal disruption by a 1000 $M_\odot$ IMBH is $2.3 \cdot 10^{-7}$ pc
(Kochanek 1992, eq. 3.2).
Since the half-mass radius of a star cluster is of order 1 in $N$-body units and globular clusters have
half-mass radii of several pc,
the adopted tidal radii correspond approximately to the tidal radius of an $m=1 M_\odot$ main-sequence star 
in a globular cluster. We assumed that a star was immediately disrupted if it entered the region with $r<r_t$
and was unaffected outside this area. The mass of a tidally disrupted star was added to the mass of the 
central black hole. To extrapolate the simulation results to larger N and compare with
other results, we always use fitting formulae which explicitly
contain $r_t$ and scale them appropriately. 

So far our runs do not incorporate the effect of gravitational radiation, which should
be unimportant for the dynamical evolution of a star cluster as long as the stellar density around the black
hole and its growth are dominated by main-sequence stars.

\begin{table*}
\caption[]{Details of the performed $N$-body runs.}
\begin{tabular}{rrcrrccr}
\noalign{\smallskip}
 \multicolumn{1}{c}{No.}& \multicolumn{1}{c}{$N$}& 
\multicolumn{1}{c}{$r_t^1$} & 
  \multicolumn{1}{c}{$M_{BH\,i}/m$} & \multicolumn{1}{c}{$M_{BH\,f}/m$} &
\multicolumn{1}{c}{$T^1_{END}$} & \multicolumn{1}{c}{$k_D$} & 
  \multicolumn{1}{c}{$N_{Dis}$} \\
\noalign{\smallskip}
 1) & 80000 &  $10^{-7}$ &  266.0 &  827.0 & 3000 & 72.0 & 561 \\
 2) & 80000 &  $10^{-7}$ &  800.0 & 1388.0 & 2000 & 63.0 & 588 \\
 3) & 80000 &  $10^{-7}$ & 2660.0 & 3285.0 & 2000 & 63.1 & 625 \\
 4) & 80000 &  $10^{-7}$ & 8000.0 & 8749.0 & 2000 & 57.0 & 749 \\
\noalign{\smallskip}
 5) &  5000 &  $10^{-8}$ &  800.0 &  815.0 & 2000 & 44.1 &  15 \\
 6) & 10000 &  $10^{-8}$ &  800.0 &  823.0 & 2000 & 54.2 &  23 \\
 7) & 20000 &  $10^{-8}$ &  800.0 &  869.0 & 2000 & 65.0 &  69 \\
 8) & 80000 &  $10^{-8}$ &  800.0 &  997.0 & 2000 & 55.3 & 197 \\
\noalign{\smallskip}
 9)&  80000 &  $10^{-7}$ &  800.0 & 1388.0 & 2000 & 63.0 & 588 \\
10)&  80000 &  $10^{-8}$ &  800.0 &  997.0 & 2000 & 55.3 & 197 \\
11)&  80000 &  $10^{-9}$ &  800.0 &  851.0 & 2000 & 73.2 &  51 \\
\noalign{\smallskip}
12)&  16000 &  $10^{-7}$ & 2589.0 & 2735.0 & 2000 & 75.6 & 146 \\ 
13)&  20000 &  $10^{-7}$ &  200.0 &  338.0 & 3000 & 53.9 & 138 \\ 
14)&  35700 &  $10^{-7}$ & 1438.0 & 1705.0 & 2000 & 63.0 & 267 \\
15)&  65536 &  $10^{-8}$ & 3276.0 & 3513.0 & 3000 & 76.9 & 237 \\
16)& 178800 &  $10^{-7}$ &  461.0 & 1368.0 & 2000 & 58.2 & 907 \\
\end{tabular}
\begin{flushleft}
Notes:  \\

1. $r_t$ and $T_{END}$ are given in
 $N$-body units.\\
\end{flushleft}
\end{table*}

In this paper, all clusters start from an initial density profile given by a King model
with $W_0=10.0$ and are composed out of equal mass stars.  Table 1 summarises
other parameters. It first shows a number identifying the run and
the number of cluster stars $N$.
Shown next are the initial and final mass of the black hole divided by the mass of a single star,
the tidal radius of the black hole, and the duration of the simulation.
The last two
quantities are given in $N$-body units in which the constant of gravitation, total cluster mass
and total energy are given by $G=1, M_C=1, E=-0.25$ respectively \citep{HeggieMatthieu1986}.
The final columns contain a dimensionless constant describing the
tidal disruption rate of stars and the total number of tidal disruptions. 
We organised the runs into 4 groups.
In the first set we varied the black hole mass and kept all other
parameters constant. The next two groups contain runs where the number of cluster stars and 
the tidal radius was varied.
The final group contains a few additional runs used in the paper.

\section{Theory}

Stars in the innermost cusp around the black hole, where the gravitational influence of the
black hole is dominating, move on 
essentially keplerian orbits with slight perturbations when they encounter other cluster
stars.
\citet{BahcallWolf1976} assumed that stars in the sphere of influence of the black hole follow
an isotropic distribution in velocity space and are absorbed if their energy becomes equal to the 
potential energy at the tidal radius. They then showed by Fokker-Planck calculations 
that the stellar density distribution 
follows a power-law $\rho(r) \sim r^{-\alpha}$ with $\alpha=7/4$ inside the sphere of influence
of the black hole down to the radius where the tidal disruption of stars becomes important.
The cusp profile will extend out to a radius where the self-gravity of the stellar system
cannot be neglected any more. If the black hole mass is  
much smaller than the mass of the stars in the core, this happens when
the velocity dispersion in the cluster
core becomes comparable to the circular velocity of stars in the field of the black hole:
\begin{equation}
 r_i = \frac{G M_{BH}}{\gamma \, <\!v_c^2\!>}
 \label{eqri}
\end{equation}
We found that $\gamma=2$ gives a good fit to the results of our simulations. The velocity dispersion of stars 
in the cluster core $<v_c^2>$ can be
expressed in terms of the core density and radius as:
\begin{equation}
 <v_c^2> = \frac{4 \pi \, G \, m \, n_c \, r_c^3}{3 \, r_c} = \frac{4 \pi}{3} \, G \, m \, n_c \, r_c^2
\label{vel}
\end{equation}
Hence the influence radius is given by:
\begin{eqnarray}
 \nonumber r_i & = & \frac{3 \, M_{BH}}{8\, \pi \, m \, n_c \, r_c^2} \\
   & \approx & 15 \, r_c \, \frac{M_{BH}}{m \, N_C} \;\;\; ,
\label{ri}
\end{eqnarray}
where $N_C$ is the number of cluster stars and we assume that the core contains roughly 3\% of
all stars in the cluster. Assuming that the cusp profile goes over into a constant density core 
with density $n_c$ at $r_i$, the number of stars in the cusp can be estimated to be
\begin{eqnarray}
 \nonumber N_{Cusp} & = & 4 \pi \int_0^{r_i} n_c \left(\frac{r_i}{r}\right)^{1.75} r^2 dr \\
  & = & 250 \; \frac{M_{BH}^3}{M_{Cl}^3} \; N_C
\end{eqnarray}
which serves to give an order of magnitude
estimate. For central black holes containing less than a percent of the total cluster mass, 
the central cusp itself contains only of order 10\% of the black hole mass in stars. Typical globular clusters
would have black holes with masses of order 1000 $M_\odot$ if they follow the relation found 
by \citet{FerrareseMerritt2000} and \citet{Gebhardtetal2000} for galaxies,
making the detection of the central cusp in their density profile difficult even with HST
\citep{DrukierBailyn2003}.
In clusters in which the black hole is more massive than the cluster core, an upper limit for
$r_i$ derives from the condition that the mass in stars inside $r_i$ should be smaller than 
the mass of the central black hole, i.e. $M(<r_i) \le M_{BH}$. 

\citet{FrankRees1976} found that $r_{crit}$, the radius at which tidal disruption of stars becomes important, 
is significantly larger than the tidal radius since relaxation lets stars drift
faster in angular momentum space than in energy space. In order to account for this, they 
introduced the loss-cone, which is the area in angular momentum space containing 
all orbits with minimum distances smaller than the tidal radius $r_t$ of the black hole. At a given
radius $r$, the opening angle of the loss-cone is given by
\begin{eqnarray}
 \theta_{lc}^2 = \frac{2 r_t}{3 r}
\end{eqnarray}
for radii $r<r_i$. They then
showed that the critical radius $r_{crit}$
is approximately given by the radius where the time for stars to drift through the loss cone due to relaxation
becomes longer than the crossing time. Inside $r_{crit}$ stars cannot drift in and out of the loss cone before 
falling into the black hole, so the loss-cone is empty.
\citet{CohnKulsrud1978} and \citet{MarchantShapiro1980} performed two-dimensional Monte Carlo calculations 
of the evolution of a star cluster with a central black hole which confirmed the loss-cone concept. 
They determined disruption rates of stars and showed
that tidal disruption of stars flattens the density profile inside $r_{crit}$
but has little influence outside this radius.

Another process which can in principle be important is the wandering of the black hole. A black hole
in a stellar system is forced to move due to 3 processes: Stars bound to the black hole
force it to move around the common centre of gravity, stars escaping from the
core cause a recoil motion of the black hole and the stellar core surrounding it, and 
passing unbound stars
cause a brownian motion of the black hole in the centre of the cluster. \citet{LinTremaine1980} 
investigated 
the role of the different processes and concluded that passing unbound stars have the largest 
effect on the black hole. For a black hole in a constant density core, they estimated the wandering 
radius to be given by: 
\begin{equation}
r_{wand} = 0.9 \, r_c \, \sqrt{\frac{m}{M_{BH}}} 
\label{wandeq}
\end{equation}
The wandering radius is difficult to determine in our simulations since it hardly exceeds
the distance of the innermost stars from the black hole and is therefore within the statistical
uncertainty with which the position of the density centre can be determined. In addition, the innermost
stars are so tightly bound to the black hole that they follow the motion of the black hole due 
to passing unbound stars as long as the unbound stars pass at large enough distances. This further complicates 
the determination
of the density centre.
\citet{Chatterjeeetal2002} gave a detailed discussion of black hole wandering and 
performed $N$-body simulations which confirmed the validity of eq.\ \ref{wandeq}. The wandering of the black
hole will limit the formation of an $\alpha=1.75$ cusp to models with large enough black hole  
masses. It could also effect the capture rate of stars since the area of the loss cone is increased.

\section{Results}

\subsection{Central density profile}

\begin{figure*}[tbp!]
\plotone{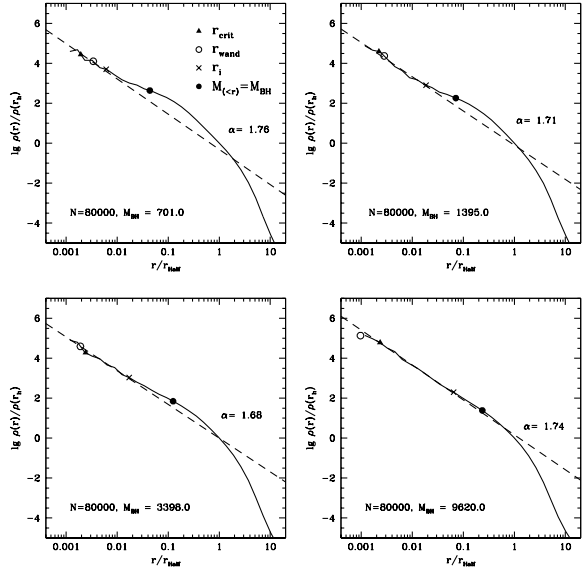}
\caption{Density profiles at $T=2000$ $N$-body units for clusters 1) to 4) from Table 1,
which contain $N=80,000$ stars and black
holes of varying masses. Solid circles mark the radii where the mass in stars becomes comparable to the
mass of the central black hole, crosses mark $r_i$ from eq.\ \ref{eqri}, open circles $r_{wand}$ and triangles
show $r_{crit}$.  An $\alpha=1.75$ power-law cusp forms inside $r_i$ in all models. 
For the model with the most massive black hole, the central cusp extends almost up to the radius 
where $M_{(<r)}=M_{BH}$.}
\label{snap_ssa}
\end{figure*}

In order to determine the central density profiles, we have overlaid 5 snapshots from the time 
the simulation was
stopped for each cluster. All snapshots were centred on the black holes.
Fig.\ \ref{snap_ssa} depicts the final density profiles inside the half-mass radius for runs 1) to
4), which contain
$N=80,000$ stars and black holes with masses between $300<M_{BH}/m<8000$. The influence radii defined
by eq.\ \ref{eqri} are marked by crosses. Also shown are the  
radii where the mass in stars becomes comparable to the mass of the central black hole (solid circle)
and the wandering radii of the black holes (open circle). For all cases studied, the critical radii are
significantly smaller than the influence radii $r_i$ and of the same order as the distance of the innermost
stars from the black hole, so we cannot test the density profile inside $r_{crit}$. The lack of stars 
inside $r_{crit}$ 
points to an efficient destruction at radii $r<r_{crit}$.

For the black hole of a few hundred solar masses, an $\alpha=1.75$ cusp cannot be found with certainty
since the number of stars inside $r_i$ is
too small. In addition, the wandering radius of the black hole is almost as large as $r_i$, so the black    
hole wandering will also flatten the profile. Clusters with more massive black holes show a clear
$\alpha=1.75$ cusp in good agreement with the prediction from Fokker-Planck and Monte Carlo models. 
The cusps extend up to radii $r=r_i$ in all cases. As expected, $r_i$ is significantly smaller than
the radius where the mass in stars becomes comparable to the mass of the central black hole unless the black
hole contains several percent of the cluster mass.  If the black hole
contains more than about 5\% of the total cluster mass, the $\alpha=1.75$ cusp goes all the way up to
the radius where the mass in stars becomes equal to the mass of the central black hole.

\begin{figure*}[htbp!]
\plotone{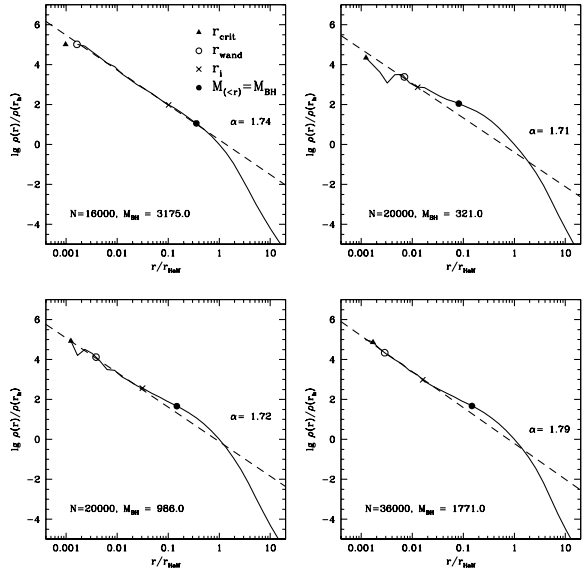}
\caption{Same as Fig.\ 1 for clusters of lower mass. All clusters have $\alpha=1.75$ slopes inside
 the radii of influence of the black hole $r_i$ and flatter slopes outside this radius.}
\label{slope} 
\end{figure*}

Fig.\ \ref{slope} depicts the central density profile for a number of small-$N$ runs. The overall behaviour
is very similar to the high-$N$ runs, showing that our results do not depend on the particle number. 
Extending our results to larger systems, we predict that in globular clusters
only of order 50 to 100 stars follow the central
cusp profile if the mass of the black hole is $M_{BH} = 1000 M_\odot$. Since only a fraction of them will
be bright enough to be easily detectable, it will be difficult to find the central cusp in the
luminosity profile of the cluster. In galactic nuclei with black hole masses in the range $10^6 M_\odot 
< M_{BH} < 10^9 M_\odot$ on the other hand, a considerable number of stars is in the $\alpha=1.75$ cusp, 
so the detection of the black hole through 
observation of the central density or velocity profile should in principle be possible.

\subsection{Accretion rates}

We can estimate the rate at which stars are disrupted by the
central black hole from the number of stars at radius $r_{crit}$ and the size of the 
loss cone, divided by the crossing time at $r_{crit}$:
\begin{equation}
\nonumber D \sim \left. \frac{r^3 \theta_{lc}^2 n(r)}{T_{CR}}\right|_{r=r_{crit}} 
\end{equation}
Since in our simulations $r_{crit}$ is much smaller than $r_i$, the density near the critical
radius follows an $\alpha=1.75$ power-law distribution:
\begin{equation}
 n(r) = n_0 r^{-1.75}
\end{equation}
Following \citet{FrankRees1976}, the critical radius can be calculated to be: 
\begin{equation}
r_{crit} = 0.2 \left( \frac{r_t \, M_{BH}^2}{m^2 \, n_0} \right)^{4/9}
\label{rc2}
\end{equation}
Hence we obtain for the disruption rate:
\begin{eqnarray}
\nonumber D & \sim & \sqrt{G} \, \left. \frac{r_t \, M_{BH}^{1/2} \,}{r^{5/4}}\right|_{r=r_{crit}} \\
   & = & k_D \, \sqrt{G} \,\frac{r_t^{4/9} \, n_0^{14/9} \, m^{10/9}}
       {M_{BH}^{11/18} }
\label{eqd}
\end{eqnarray}

\begin{figure*}[htbp!]
\plotone{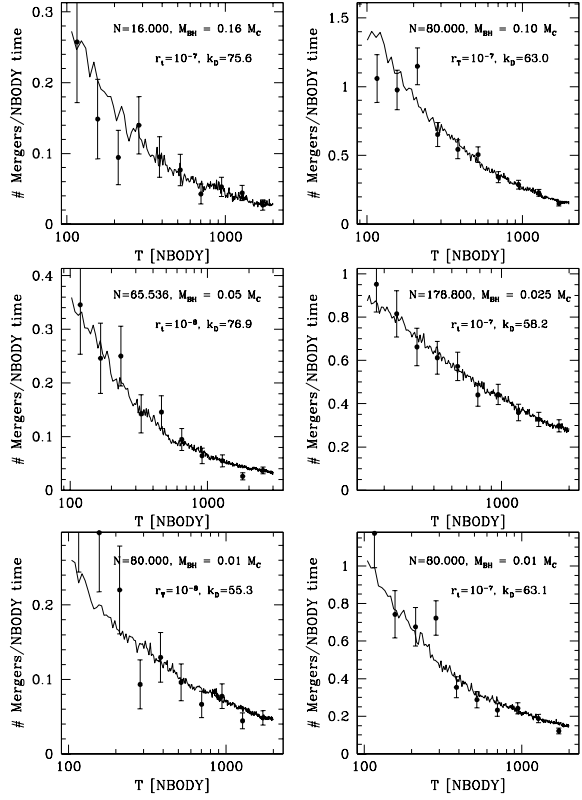}
\caption{Evolution of the tidal disruption rate of stars with time for 6 different clusters. Points with
 errorbars are results from $N$-body simulations. Solid lines show a fit according to eq.\
 \ref{eqd} with the constants $k_D$ adjusted to match each case. The tidal disruption rate drops due
  to the expansion of the clusters. There is good agreement between the $k_D$ values for the 
  different clusters.}
\label{merg_t2}
\end{figure*}

Fig.\ \ref{merg_t2} shows the evolution of the disruption rate as a function of time for 
clusters with a range of black holes masses. In all runs the tidal disruption rates
decrease due to the cluster expansion which will be discussed in the next paragraph. 
The cluster expansion decreases the stellar density near $r_{crit}$ and increases the crossing time, 
thereby decreasing $D$. Solid lines show expected disruption rates according to eq.\ \ref{eqd}, calculated by 
determing $n_0$ from the actual $N$-body data and the constants $k_D$ from a best-fit to the overall disruption rate.
The time evolution of the disruption rates calculated this way agrees very well with the one found 
in the $N$-body simulations. 
In addition, the constants $k_D$ determined for the different simulations also agree reasonably well with
each other and most are compatible with an average of $k_D=65$ (see Table 1). 

In order to calculate the collision rate in physical units, we assume that the
tidal radius of a star with radius $R_*$ and mass $m$ is given by: 
\begin{equation}
 r_t = 1.3 \, R_* \left( \frac{M_{BH}}{2 \, m} \right)^{1/3}
\end{equation}
(Kochanek 1992, eq. 3.2). We then obtain for the tidal disruption rate:
\begin{eqnarray}
\nonumber D & = & \frac{0.00588}{100 \; \mbox{Myrs}} \left(\frac{R_*}{R_\odot}\right)^{\frac{4}{9}} \! \left( \frac{n_0}
  {\mbox{pc}^{-1.25}} \right)^{\frac{14}{9}} \! \\
 & & \left(\frac{m}{M_\odot}\right)^{\frac{26}{27}} \! \left(\frac{M_{BH}}{1000 M_\odot}\right)^{-\frac{25}{54}}
\label{d1}  
\end{eqnarray}
Although eq.\ \ref{d1} seems to indicate that $D$ decreases with the black hole mass, this is not 
the case since the cusp density constant $n_0$ also depends on the black hole mass and 
the central density $n_c$. 
It is however interesting to apply eq.\ \ref{d1} to cusps of observed systems. \citet{Genzeletal2003}
for example derived a stellar density of $\rho = 1.2 \cdot 10^6 \, \mbox{M}_\odot \mbox{pc}^{-3}$ at a distance of $r=0.38$ pc 
from the 
galactic centre. They found that the density inside this radius rises with a power-law with power $\alpha=1.4$,
while it falls off with $\alpha=2.0$ outside this radius. Both values are not too far from the $\alpha=1.75$ slope
in our runs. Using their density, we find $n_0 = 2.21 \cdot 10^5 \mbox{pc}^{-1.25}$ if we assume an average stellar mass of
$<\!m\!>=1 \mbox{M}_\odot$. From eqs.\ \ref{rc2} and \ref{d1} we then obtain a critical radius of 
$r_{crit}=1.2$ pc and a total number of disruptions of $D=30,000$ per $T=100$ Myrs for a central black hole
mass of $M_{BH}=3 \cdot 10^6 M_\odot$. Tidal disruption of stars could therefore play an important
role for the growth of the galactic centre black hole.

In order to apply eq.\ \ref{d1} to systems with a constant density core, we assume that the
cusp density goes over into a constant density core with density $n_c$ at $r=2 r_i$. With the help of eq. \ref{ri}
we then obtain
\begin{eqnarray}
\nonumber D &\!\!=\!\!& \frac{17504.9}{100\, \mbox{Myrs}} \, \left( \frac{R_*}{R_\odot} \right)^{4/9} 
  \left(\frac{m}{M_\odot} \right)^{-95/54} \\
 & & \left( \frac{n_c}{\mbox{pc}^{-3}} \right)^{-7/6}\!\! \left( \frac{r_c}{pc} \right)^{-49/9} \!\!
   \left(\frac{M_{BH}}{1000 M_\odot} \right)^{61/27} 
\label{d2} 
\end{eqnarray}
The powers of the different factors are the same as the ones obtained by Frank \& Rees for the case
$r_{crit}<r_i$ (eq.\ 16a in their paper). Our disruption rate is about a factor of 2 larger than theirs.
From eq.\ \ref{vel}, we can obtain an
alternative expression which uses the core velocity dispersion, which is easier to observe than the
core radius: 
\begin{eqnarray}
\nonumber D&=&\!\!\frac{22.9}{100\, \mbox{Myrs}}  \left( \frac{R_*}{R_\odot} \right)^{4/9}
 \left(\frac{m}{M_\odot} \right)^{26/27}\!\! \left( \frac{n_c}{5\cdot10^4 \, \mbox{pc}^{-3}} \right)^{14/9}\!\! \\
  & & \cdot \left( \frac{v_c}{10 km/sec} \right)^{-49/9} \, \left(\frac{M_{BH}}{1000 M_\odot} \right)^{61/27}
\label{d3}
\end{eqnarray}
This formula agrees with the one from \citet{CohnKulsrud1978} (eq. 66),
who obtained nearly the same dependence of the disruption rate on the different physical
parameters. Our disruption rate is smaller by about $33\%$, which indicates a very good agreement given
the errors involved by assuming that the central cusp goes over directly into a constant density 
core, which is a significant simplification of the real situation (Figs.\ 1 and 2).

Typical parameters for the densest globular clusters 
are $n_c=5\cdot10^5$/pc$^3$ and $v_c=15$ km/sec. While a $M_{BH}=1000$ black hole should be able to double
its mass within a few Gyrs from disrupted main-sequence stars in such a cluster, it does not seem
possible to grow a 1000 $M_\odot$ black hole from a 100 $M_\odot$ progenitor within a Hubble
time. A more detailed discussion must however also take the mass distribution of the stars into
account, which will be the focus of a follow up paper.

\begin{figure}[tb!]
\epsscale{1.0}
\plotone{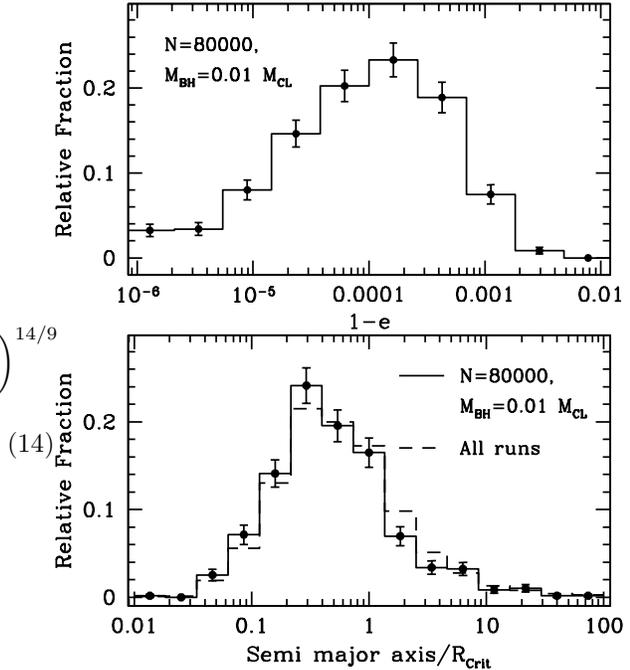}
\caption{Distribution of orbital eccentricities (above) and semi major axis (below) of stars
 being tidally disrupted by the central black hole for run 2) which has $N=80,000$ and $M_{BH} =
 0.01 M_{Cl}$. All disrupted stars move on highly eccentric orbits,
  implying that drift in angular momentum space is the main process contributing to the tidal disruption 
   of stars. Stars being tidally disrupted by the black hole originate from approximately the critical 
    radius (bottom panel).} 
\label{mergd}
\end{figure}

Fig.\ \ref{mergd} shows the eccentricity distribution of stars which are tidally disrupted
by the black hole and their semi major axis $a$ in relation to the critical radius.
Stars that are disrupted by
the central black hole move on very eccentric orbits on average, with practically all of them 
having orbital eccentricities $e > 0.999$ on the final orbit prior to disruption, 
in good agreement with the idea that the drift in angular momentum space is 
more important than the drift in energy space for the feeding of the black hole. 
The lower panel shows the semi major axis distribution.
Most stars disrupted by the black hole have semi major axis $a \le r_{crit}$. The median $a$ turns 
out to be about
$a = 0.5 r_{crit}$. Since the eccentricity is nearly unity, the apobothron distance, i.e. the maximum 
black hole distance 
on the last orbit before disruption is $r_a=2 a$, so $r_a \approx r_{crit}$ in good agreement
with our numerical estimates for $r_{crit}$.
We do not find strong evidence for significant
differences in the distribution of $a/r_{crit}$ in the different runs, so we can average them. The dashed curve in the
lower panel shows the distribution averaged over all runs, which is nearly identical to the
solid curve for one particular run.
Since in our runs $r_{crit}<<r_i$ all stars being disrupted by the black hole are tightly bound
to it. As these stars are
constantly replenished by less bound stars from outside $r_i$, 
energy conservation requires that the rest of the system gains energy and
expands. This expansion will be the focus of the next section.

\subsection{Cluster evolution}

\begin{figure}[tb!]
\plotone{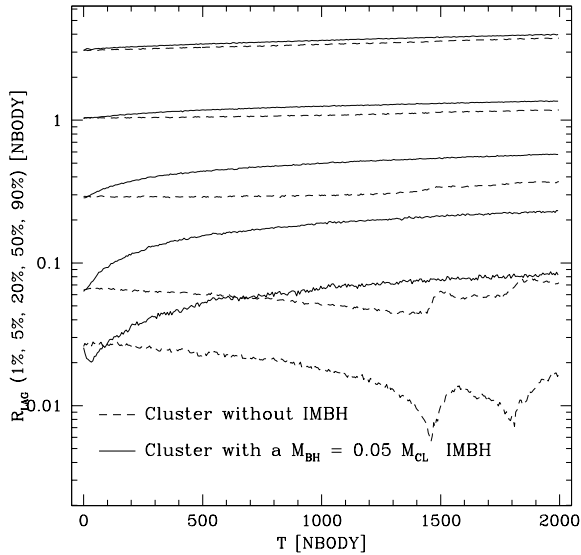}
\caption{Evolution of lagrangian radii for two $N=65536$ star clusters, one with a massive black hole
and one without. The cluster without black hole goes into core-collapse at $T=1450$ $N$-body units and expands
during the post-collapse phase due to energy generated by binaries formed in the core.
The cluster with a black hole expands from the start since energy exchanges between stars in the 
 cusp around  
the black hole provide the energy for the expansion.}
\label{exp1}
\end{figure}

\begin{figure}[tb!]
\plotone{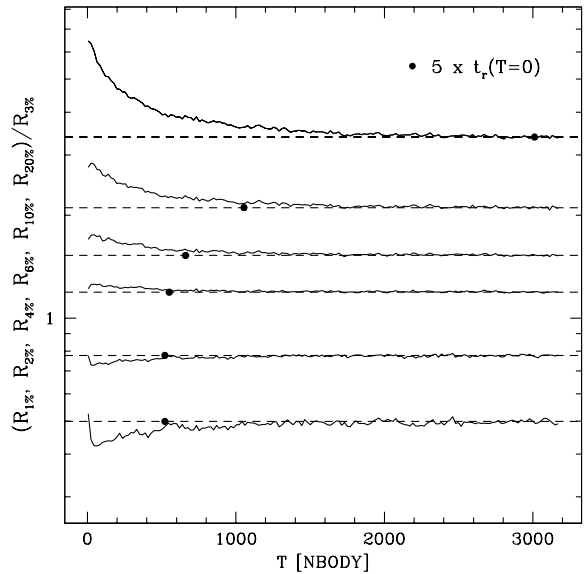}
\caption{Ratio of the lagrangian radii to the 3\% lagrangian radius for the run with a 1000 $M_\odot$
black hole from Fig.\ \ref{exp1} (solid lines). Dashed lines show average ratio near the end of the simulation. After
an initial phase where the core is re-adjusting itself,
the ratios of different radii become constant and the cluster expansion becomes self-similar. The black
dots show the time when 5 initial relaxation times have passed at the different radii.}
\label{exp2}
\end{figure}

The evolution of lagrangian radii of two clusters is depicted in Fig.\ \ref{exp1}. Shown is cluster 
15), which contains
$N=65536$ 
stars and a black hole of $M_{BH} = 0.05 M_{Cl}$ and a second cluster with the same initial density profile and
number of cluster stars but no massive black hole.  The cluster without the
black hole goes into core-collapse at $T=1460.0$ $N$-body units.
Core-collapse is halted when binaries form in the core and interactions between the binaries
and passing cluster stars heat the cluster and lead to an overall expansion of the cluster. Since the
cluster is isolated and we did not allow for stellar collisions,
no characteristic length scale exists and the expansion 
is proportional to 
\begin{equation}
r \sim t^{2/3} 
\end{equation}
in the post-collapse phase \citep{GierszHeggie1994, Baumgardtetal2002}.
The cluster core undergoes gravothermal oscillations since the number of active binaries in the core is small: 
A look at the cluster data shows that most of the time there is only one binary in the core which powers the expansion,
in agreement with theoretical expectations \citep{Goodman1984}. If this binary is expelled,
the cluster centre re-collapses again and forms new binaries. In contrast, 
the cluster with a central black hole expands right from the start and without core-oscillations
since the expansion is powered by energy exchanges of stars in the central cusp around the
black hole, which remains in the cluster core due to its high mass. Only the innermost radii show
a short collapse phase in the beginning, when the central cusp profile is created. Binaries 
cannot be responsible for the heating since their formation is suppressed by the high stellar
velocities in the cusp and the strong gravitational 
field of the black hole which disrupts binaries. Consequently, no stable systems formed in our 
runs. Fig.\ \ref{exp1}
confirms that a central black hole can act as a heat source similar to the effect of 
binaries in post-collapse clusters. 

Fig.\ \ref{exp2} shows the evolution of lagrangian radii for cluster 15), which has $N=65536$ 
stars and
a black hole mass of $0.05 M_{Cl}$. All radii are divided by the 3\% lagrangian radius. If we calculate
the local relaxation time at different lagrangian radii from \citet{Spitzer1987}:
\begin{equation}
 t_{r} = 0.065 \; \frac{v_m^3}{n m^2 ln \Lambda}
\end{equation}
with $\Lambda= 0.11 N$ \citep{GierszHeggie1994}, it
can be seen that the ratio of the different radii to the 3\% radius becomes constant after
about 5 local relaxation times have passed. The expansion of the cluster becomes therefore self-similar
beyond this time. Since the relaxation time
increases with radius, the equilibrium profile is established first for radii nearest to the black
hole and then forms at larger radii. We obtain similar results for the other clusters.
Most globular clusters have central relaxation times much less than a Hubble time (Spitzer1987, Fig 1.3), 
so we expect that they
have reached equilibrium profiles in their centres if they contain massive black holes.

\begin{figure}[tb!]
\plotone{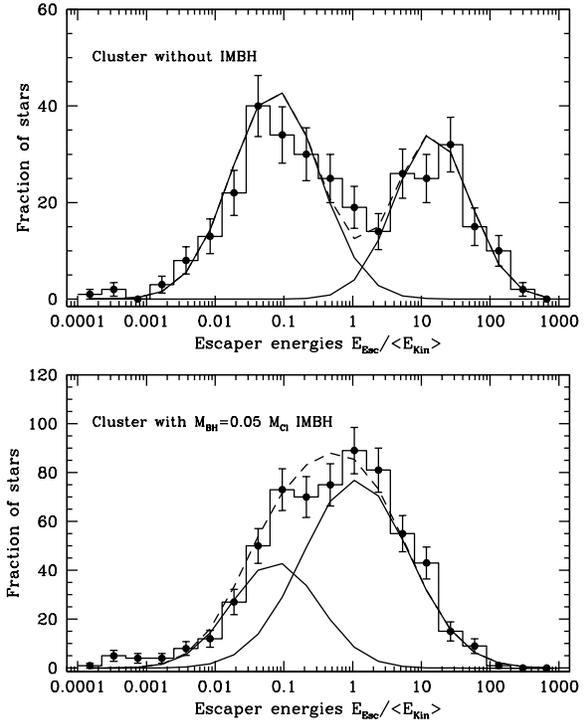}
\caption{Distribution of escape energies for the clusters of Fig.\ \ref{exp1}. For a cluster
without central black hole, two distinct peaks are visible, corresponding to stars escaping due
to distant two-body encounters and close three-body encounters involving binaries. In the cluster with
a black hole, most stars escape due to close encounters in the cusp around the black hole. Solid
lines are gaussian fits to the energy distributions of different
escape mechanisms (see text). Dashed lines show the sums, which fit the $N$-body results very well.}
\label{esc_ene}
\end{figure}

Closely related to the expansion of a cluster is the escape of stars.
The energy distribution of stars escaping from the two clusters of Fig.\ \ref{exp1} is depicted in 
Fig.\ \ref{esc_ene}. 
We measure the energies of escaping stars after they have left the cluster and divide the energies by the average
kinetic energy of all stars still bound to the cluster at the time the escape event happens. The energy
distribution of escapers in case of no black hole shows two distinct maxima,
corresponding to escapers created by a slow diffusion process in the outer parts of the cluster 
and escapers created by three-body encounters in the cluster core. We can fit the whole escaper distribution by the
sum of two gaussians with maxima at $<\log{E/E_{Kin}}> = 0.1$ and 10.0, similar to what \citet{Baumgardtetal2002}
found for isolated clusters with lower $N$. 

Although the structure of both clusters is
not exactly the same and a direct comparison therefore difficult, it can be seen that the number of escapers 
is increased if a black hole is present (note different scale on y-axis in lower panel). 
At T=1450, at which time the cluster without IMBH reached core collapse, the cluster with a black hole has
created about 6.5 times as many escapers as the cluster without black hole. In the post-collapse phase, the
number of escapers per time is approximately equal in the two runs. Since the process for generating escapers 
is different, 
the energy distribution of escapers in the two runs is also different especially at the high-E end. Escapers
created by distant encounters outside the core should still be present in the black hole case. If we subtract the  
distribution found for stars escaping due to distant encounters in case of no black hole from the black
hole case, we can fit
the remaining distribution by a gaussian distribution with mean $<\log{E/E_{Kin}}> = 1.0$. These are escapers 
created in the high-density cusp around the black hole. From Fig.\ 
\ref{esc_ene} we can estimate that about 
200 stars escape by distant encounters and about 420 by close encounters in the cusp. During the same time
211 stars are disrupted by the black hole, so stars in the cusp have a larger chance of escape than 
disruption. Nevertheless, the energy carried away by the escapers is small compared to the energy created 
by the tidal disruption of stars, since we find that about 10 times as much energy is created than taken
away by escapers. The energy created by the tidal disruption of stars therefore enters mostly the cluster
expansion. We obtain this result in all simulated clusters.

Mathematically, when a star is accreted to the central BH, the binding
energy of the star cluster is reduced by the amount same as the
binding energy of the star and the IMBH. In other words, the cluster is
heated up. Since almost all stars accreted to BH are strongly bound to
IMBH, accretion almost always resulted in heating. From a physical point
of view, accreted stars increased their binding energy by giving their
kinetic energies to field stars, thus heating the cluster. How much
heat the accreted star gave is simply determined by its binding energy
at the moment of the accretion.
The effective energy generation by the stars disrupted by the black hole can be estimated from the
disruption rate multiplied by the typical energy of a star disrupted by the black hole. Since 
$r_{crit}>>r_t$,
stars disrupted by the black hole are on nearly radial orbits, so the energy of a star is equal to 
the potential energy at $r_{crit}$: $E_*=G \, M_{BH} \, m/r_{crit}$. We therefore obtain 
\begin{eqnarray}
\nonumber \dot{E} & = & D \cdot E_\star \\ 
 & = & k_D \, G^{3/2} \,\frac{m^3 \, n_0^2}{M_{BH}^{1/2}}
\end{eqnarray}
Interestingly, there is no dependence of the energy generation rate on the tidal radius $r_t$. 
For small mass black holes, most of the potential energy comes from the gravitational interaction
between the stars
themselves. The energy needed to expand the cluster is therefore given by
\begin{equation}
\dot{E} = k \frac{G \; M^2_{Cl}}{r_h^2} \;\; \dot{r}_h \;\; ,
\end{equation}
with $r_h$ being the cluster's half-mass radius and $k$ a constant of order unity. Combining both
equations and assuming that the cusp profile goes over into a constant density core at $r_i$, we
obtain
\begin{equation}
 \dot{r}_h \sqrt{r_h} = \sqrt{G} \, 30 \, k_D \left( \frac{r_h}{r_c} \right)^{2.5} \frac{M^3_{BH} \, m}{M_{Cl}^{3.5}}
\end{equation}
With $k_D=65$, $M_{Cl}=1.0$ and $r_h/r_c \approx 10$ this gives
\begin{equation}
r_h^{3/2} - r_{h0}^{3/2} = 6 \cdot 10^5 \, M_{BH}^3 \, m \, t
\end{equation}
The time dependence is the same as in the self-similar case, $r \sim t^{2/3}$, since the energy generation
rate is independent of the tidal radius, so no characteristic length scale exists.
\begin{figure}[tb!]
\plotone{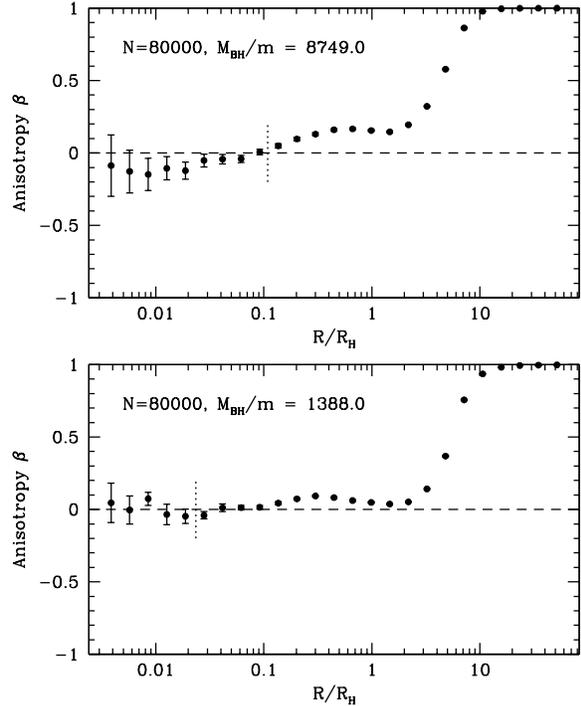}
\caption{Anisotropy as a function of radius for two clusters with $N=80,000$ stars. The
dotted vertical line marks the radius of influence of the black hole.
The cluster with the larger black hole has a slight tangential anisotropy in the
centre due to the preferential disruption of stars on radial orbits.}
\label{aniso}
\end{figure}

Fig.\ \ref{aniso} depicts the velocity anisotropy profile for runs 2) and 4), which have
$N=80,000$ clusters and black holes
containing 1\% and 10\% of the total cluster mass. The anisotropies were defined as
\begin{equation}
 \beta = 1-\frac{\sum_i v^2_t}{2 \sum_i v^2_r} 
\end{equation}
where $v_r$ and $v_t$ are the radial and tangential velocities of each star and the sum runs over
all stars in a radial bin.
No significant degree of anisotropy can be detected in the cluster with the small black hole
inside the influence radius of the black hole.
The cluster with the more massive black hole has a slightly tangentially anisotropic
velocity dispersion, similar to what \citet{CohnKulsrud1978} found 
in their Fokker-Planck calculations for stars inside the influence radius of the black hole.
The reason is the preferential
disruption of stars on radial orbits and the fact that more stars are disrupted in a cluster
with a more massive black hole.
\citet{Amaro-Seoaneetal2004} (their Fig.\ 8) obtained a radially anisotropic profile between
$r_{crit}<r<r_i$, which could be due to their very large half-mass radius at the end of the run,  
which decreases the merging rate of stars. In any case, the small amount of anisotropy
together with the small number of stars inside $r=r_i$ makes an observation of this effect 
for globular clusters almost impossible, although it might be detectable in galactic nuclei.
For isolated clusters, cluster halos are build up from stars scattered out of the
centre which move on very radial orbits. Therefore, both clusters have radially anisotropic profiles
outside the half-mass radius. For any realistic
cluster, the tidal field would cause an isotropisation of the stellar orbits.

\section{Conclusions}

We report the first results of self-consistent $N$-body simulations
of star clusters composed of equal-mass stars and a central massive black hole.
We find that in clusters with a massive central black hole a $\rho \propto r^{-1.75}$ power-law
cusp forms inside the sphere of influence of the black hole, in good agreement with predictions from 
Fokker-Planck and Monte Carlo simulations. In star clusters where the black hole mass is less than a
few percent of the total cluster mass, the cusp contains only a fraction of the black hole mass in stars.
The minimum black hole mass needed to form an $\alpha=1.75$ cusp
is several $100 M_\odot$ for a typical globular cluster.
Otherwise the cusp contains too few stars to be significant. For a black hole mass 
less than a few percent of the total cluster mass, the density profile
is shallower than $\alpha=1.75$ outside the radius of influence $r_i$ of the black hole. For more massive 
black holes, the $\alpha=1.75$ cusp extends all the way through the core.

The cusp profile forms from the inside out and it takes about 5 local relaxation
times until it is established at a given radius. Since central relaxation times of globular clusters are 
of order $10^7$ to $10^8$ years, globular clusters should have $\alpha=1.75$ power-law cusps
in their centres and will evolve more or less along a sequence of equilibrium profiles if they
contain massive black holes.
Inside the radius of influence of the black hole, the velocity profile is slightly tangentially anisotropic
due to the tidal disruption of stars on radial orbits. Since the magnitude of this effect is small,
it is unlikely that it will be detectable except for star clusters with very massive
black holes.

Our simulations confirm the merging rates found in Fokker-Planck simulations and from analytic estimates.
Tidal disruption of stars could play an important role for the current growth of the super-massive black hole
at the galactic centre and the growth of intermediate-mass black holes in dense star clusters. 
However, the
formation of an intermediate-mass black hole out of a stellar mass one in a globular cluster by tidal 
disruption of stars alone seems impossible.
A massive black hole in a star cluster merges
mainly with tightly bound stars from its direct vicinity. These stars are constantly replaced by less bound 
cluster stars which drift inward due to relaxation. This causes an overall expansion of the cluster. Black 
holes can therefore halt
the core collapse of globular clusters, similar to the effect of binaries in post-collapse clusters.
As in the case of an expansion driven by a central population of binaries, the expansion is self-similar, 
i.e. $r \sim t^{2/3}$.

\section*{Acknowledgements}
We are grateful to Sverre Aarseth for his constant help with the NBODY4 code. We also thank
the referee, Fred Rasio, for a careful reading of the manuscript and useful comments.
H.B. is supported by the Japan Society for the Promotion of Science through Grant-in-Aid for JSPS 
fellows 13-01753.

\end{document}